\documentclass[11pt,journal,comsoc]{IEEEtran}

\usepackage[T1]{fontenc}
\usepackage[noadjust]{cite}

\usepackage[cmintegrals]{newtxmath}
\usepackage[obeyspaces]{url}
\usepackage{cite}
\usepackage{amsmath}
\usepackage{algorithmic}
\usepackage{graphicx}
\DeclareGraphicsExtensions{.pdf,.jpeg,.png,.svg,.pdf_tex}
\usepackage{tabularx}
\usepackage{tabulary}
\usepackage{textcomp}
\usepackage{xcolor}
\usepackage{fancyhdr}

\def\BibTeX{{\rm B\kern-.05em{\sc i\kern-.025em b}\kern-.08em
    T\kern-.1667em\lower.7ex\hbox{E}\kern-.125emX}}
\newcommand{\REV}[1]{{\color{black} #1}}

\begin{document}

\title{Holistic Privacy for Electricity, Water, and Natural Gas Metering in Next Generation Smart Homes}
%\title{Holistic Privacy Protection for Electricity, Water, and Natural Gas Metering}

\author{
Cihan~Emre~Kement,~\IEEEmembership{Student~Member,~IEEE}, Bulent~Tavli,~\IEEEmembership{Senior~Member,~IEEE}, Hakan~Gultekin, and Halim~Yanikomeroglu,~\IEEEmembership{Fellow,~IEEE}

\thanks{\textit{Cihan Emre Kement is with the Department of Electrical and Electronics Engineering, TOBB University of Economics and Technology, and Laboratory for Information and Decision Systems, Massachusetts Institute of Technology (e-mail: kement@mit.edu).}}
\thanks{\textit{Bulent Tavli is with the Department of Electrical and Electronics Engineering, TOBB University of
Economics and Technology (e-mail: btavli@etu.edu.tr).}}
\thanks{\textit{Hakan Gultekin is with the Department of Mechanical and Industrial Engineering, Sultan Qaboos University, and the Department of Industrial Engineering, TOBB University of Economics and Technology (e-mail: hgultekin@squ.edu.om).}}
\thanks{\textit{Halim Yanikomeroglu is with the Department of Systems and Computer Engineering, Carleton
University (e-mail: halim@sce.carleton.ca).}}}

\maketitle

\thispagestyle{fancy} 
\chead[H]{This work has been submitted to the IEEE for possible publication. Copyright may be transferred without notice, after which this version may no longer be accessible.}

\begin{abstract}
\label{section:abstract}
In smart electricity grids, high time granularity (HTG) power consumption data can be decomposed into individual appliance load signatures via Nonintrusive Appliance Load Monitoring techniques
%, which can be exploited
to expose appliance usage profiles. Various methods ranging from load shaping to noise addition and data aggregation have been proposed to mitigate this problem. However, with the growing scarcity of natural resources, utilities other than electricity (such as water and natural gas) have also begun to be subject to HTG metering, which creates privacy issues similar to that of electricity. Therefore, employing privacy protection \REV{countermeasures} for only electricity usage is ineffective for appliances that utilize additional/other metered resources. As such, existing privacy \REV{countermeasures} and metrics need to be reevaluated to address not only electricity, but also any other resource that is metered. Furthermore, a holistic privacy protection approach for all metered resources must be adopted as the information leak from any of the resources has a potential to render the privacy preserving \REV{countermeasures} for all the other resources futile. This paper introduces the privacy preservation problem for multiple HTG metered resources and explores potential solutions for its mitigation.

\begin{IEEEkeywords}
IoT, nonintrusive appliance load monitoring, privacy, smart metering, smart natural gas metering, smart water metering
\end{IEEEkeywords}
\end{abstract} 

\section{Introduction}
\label{section:introduction}
\IEEEPARstart{S}{mart} Grid (SG) concept has emerged as the traditional power grid could no longer address the contemporary needs of utility companies (UCs) and consumers. Smart metering is one of the fundamental enabling technologies of SG which describes the deployment of electric meters and associated infrastructure facilitating two-way communications between meters and the central system. Indeed, smart metering is an integral part of the Internet-of-Things (IoT).

\begin{figure}[hbt]
    \centering
    \vspace{-0.5cm}
    \begin{center}
        \includegraphics[width=0.5\textwidth]{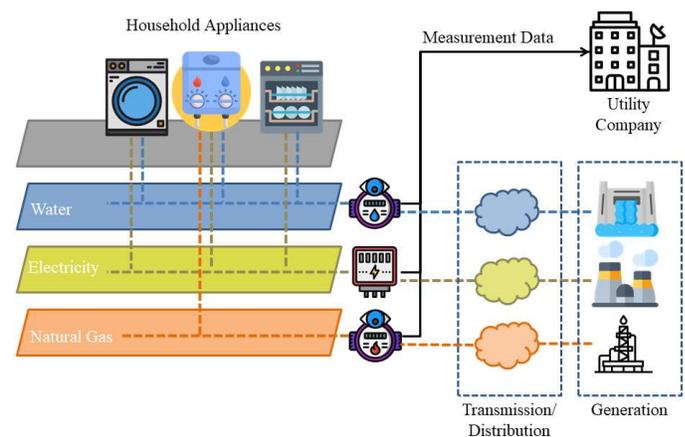}
        \vspace{-0.5cm}
        \caption{Residential smart electricity, water, and gas metering.}
        \label{figure:smart}
    \end{center}
    \vspace{-0.5cm}
\end{figure}

%Thanks to High Time-Granularity (HTG) smart metering, electricity usage of consumers can be metered sub-hourly (even per minute) and this data can be used to better manage the grid and also to provide accurate and personalized billing services. 
High time granularity (HTG) smart metering enables minutely metering of the electricity usage of consumers and this data can be used to better manage the grid and also to provide accurate and personalized billing services. However, the frequent metering of electricity usage brought along privacy concerns, since there are a variety of Nonintrusive Appliance Load Monitoring (NIALM) methods which can be employed to infer individual appliance usages from the metered energy utilization of the consumer~\cite{marmol2012do}. As the natural resources are becoming more and more scarce, within the umbrella term of next generation smart home, resources other than electricity have also become subject to HTG smart metering (see Fig.~\ref{figure:smart}). 
The total number of HTG electricity, gas, and water meters is expected to continue surging (more than \emph{200 million} % smart meters are expected
to be deployed only in 2024~\cite{iot}) along with the privacy concerns associated with it.
%The total size of the smart electricity, gas, and water meter market is expected to continue surging, surpassing \emph{26 billion dollars} by 2025~\cite{mam}, along with the privacy concerns associated with it.

The term NIALM has emerged in 1980s and since then it has been \REV{a generic descriptor} for methods and techniques that identify appliance load signatures from the aggregate metered load data of a residential (e.g., a house) or commercial (e.g., a factory) electricity consumer. Since these methods can infer individual appliance loads from the overall metered load of a consumer, there is no need to put sensors or meters inside the house/workplace, hence, the term "nonintrusive" is being used. To render NIALM methods ineffective, many \REV{countermeasures} have been proposed in the literature, each having their own advantages and shortcomings. These methods can be broadly categorized into five: data obfuscation, aggregation, anonymization, downsampling, and load shaping, which are further explained later in this article.

Each of these methods adopts a different approach to address the privacy preservation in SGs to counter NIALM against electricity utilization. However, recent developments in HTG metering of other resources such as water and natural gas bring their own privacy issues. Furthermore, information leakage from any resource can undermine the effectiveness of the \REV{countermeasures} adopted for other resources.

\begin{figure}[hbt]
	\centering
    \vspace{-0.5cm}
    \begin{center}
        %\shorthandoff{=}
        \includegraphics[width=0.5\textwidth]{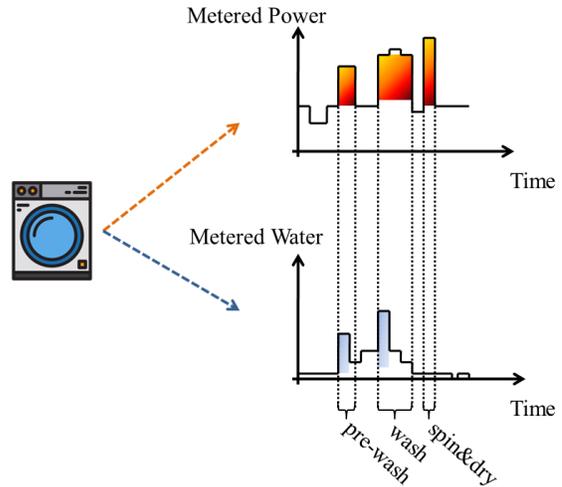}
        \caption{Appliances that use multiple resources (such as washing machines) leave traces on multiple smart meters.}
        \label{figure:signatures}
    \end{center}
    \vspace{-0.5cm}
\end{figure}

The literature on privacy threats and \REV{countermeasures} against them within the context of smart metering is limited to electricity metering only. To the best of our knowledge, there is no study that focuses on privacy issues in smart metering of water and gas. Furthermore, we are not aware of any previous investigation on holistic privacy preservation for electricity, gas, and water metering. One of the reasons for such a gap is that HTG water and gas metering are rather recent trends unlike HTG electricity metering which is more widespread and rather a mature technology. Nevertheless, there is a growing issue of privacy for water and natural gas metering which will not only pose a serious problem just as SG privacy did for a while now, but also renders current techniques ineffective because hiding only electricity usage will not be enough to hide the usage of appliances that use electricity and, at least, one of the other resources (see Fig.~\ref{figure:signatures}). This study explores the emerging problem of privacy preservation for the next generation smart homes with multiple smart metered resources. Our novel contributions are summarized as follows:
\begin{itemize}
  \item We justify the need for privacy protection in water and gas utilization by reviewing deployments of HTG smart water and gas metering and NIALM techniques designed specifically for them.
  \item We review the existing privacy protection techniques for HTG electricity metering and demonstrate how these methods can be adapted to HTG water and gas metering.
  \item We introduce the concept of holistic privacy for smart homes utilizing HTG smart metered electricity, gas, and water.
\end{itemize}

Next section overviews smart metering and NIALM techniques utilized for metered water and gas. In the subsequent sections; adaptation of privacy preserving methods employed in SGs to metered water and natural gas, and the concept of holistic privacy is introduced and elaborated. Concluding remarks and future research directions are provided in the last section of the article.

\section{Smart Resource Metering}
\label{section:advances}

Utilization of resources such as electricity, water, and gas has been increasing in proportion to the growing global economy. As these resources are inherently limited, efficient consumption of them has become a priority for sustainable growth. Smart metering is used to facilitate this efficiency. Smart electricity metering and its privacy issues have been studied widely in the literature~\cite{sun2015comprehensive,iot,kement2017comparative,marmol2012do}. However, there are also significant advancements on smart water and gas meters with high precision and high frequency metering capabilities. We overview smart metering technology, deployment status, and NIALM for gas and water in the rest of this section. 

\subsection{Smart Metering Technology}
\label{subsec:smart-metering-technology}

Traditional water meters have low measurement precision and low time granularity when compared to the modern smart metering. Indeed, typically, water consumption is measured in terms of m$^3$ or ft$^3$, while the most frequent readings are done manually once in a month, coinciding with the billing period. Currently, water meters with built-in Advanced Metering Infrastructure (AMI) support for automatic collection of water usage data are available. This gives UCs the opportunity to monitor water usage much more frequently.

While it is possible for the UCs to shorten the reading period arbitrarily using current smart water meters, increasing the measurement precision requires adoption of alternative technologies. There are several sensors that promise various sampling frequencies and metering resolutions~\cite{mudumbe2015smart,cominola2015benefits,schantz2014water}. Flow meters support a metering frequency up to 1 Hz and resolution of 0.014 liters. Pressure sensors can have a metering frequency of 2 kHz. Accelerometers can measure the water usage up to 100 Hz and have a resolution of 0.015 liters. The most precise metering can be achieved by using ultrasonic sensors with sampling frequency of up to 2 MHz and a resolution of \REV{1.8~mL}.

Measuring gas is inherently more difficult than measuring water, because its volume depends on the instantaneous temperature and pressure. Nonetheless, there are several studies regarding HTG smart gas metering~\cite{tewolde2010high, dong2017mems}.

Traditional gas meters such as diaphragm and rotary meters use positive displacement for measurement and have a resolution of a few ft$^3$ with accuracy up to 0.25-1\%. By utilizing a special setup with a high resolution encoder, their resolution can be increased to 0.005 ft$^3$~\cite{tewolde2010high}. Velocity meters such as turbine gas meters can perform measurements up to 1 kHz and provide accuracy of 0.5-1\%. Differential pressure meters such as orifice gas meters can have the accuracy of 0.25-2\%. Other technologies such as ultrasonic flow meters and Coriolis meters promise even higher accuracy levels~\cite{sun2015comprehensive}.

\REV{NIALM methods are effective when the metering is frequent enough to capture the ON/OFF events of the appliances. 
%For example, if metering is done every 10 minutes, an appliance whose runtime is 5 minutes may not be detected from the measurements. 
For example, if metering is done every 10 minutes then an appliance with a runtime of 5 minutes can, potentially, be invisible to NIALM methods.
However, aforementioned water and natural gas sensors have metering frequencies that are higher than 1 Hz. Therefore, it is possible to capture the ON/OFF events of all conventional water/gas end-uses/appliances. Similarly, for NIALM methods to detect an appliance, the resolution of the meters should be smaller than the the appliance's consumption during one metering period. For example, if water is measured every minute and an appliance uses 1 liter of water in a minute, the water meter's resolution should be less than 1 liter in order to accurately detect the ON/OFF event of the appliance. Nevertheless, resolutions of the state of the art water and gas sensors are more than enough for detecting domestic end-uses/appliances.}

\subsection{Smart Meter Deployment}
\label{subsec:smart-meter-deployment}
Although overlooked by many due to their slow progress compared to the deployment of smart electricity meters, smart water and natural gas meters are being deployed all around the world, especially in developed countries and areas. The increasing scarcity of natural resources gives rise to the speed of the progress.
%Currently, more than 50 countries are in the progress of deploying smart meters~\cite{sun2015comprehensive}.

Smart natural gas meters are replacing the legacy ones in most of the developed countries and the emerging economies. The UK recently started to install smart gas meters to reach, eventually, more than 30 million residences. Italy plans to deploy smart gas meters to 100\% of the households by the end of 2020~\cite{nhede_2018}.

Smart water meters, on the other hand, are being deployed throughout the globe where financing such a great investment is possible and, especially, in the countries where clean water resources are scarce, such as Australia, New Zealand, India, and South Africa among others. The number of smart water meters installed worldwide is expected to reach 400 million by 2026~\cite{cominola2015benefits}.

\subsection{Nonintrusive Appliance Monitoring}
\label{subsec:niam}

High granularity water metering data facilitates the disaggregation of a household's total water usage into individual appliances or end-use categories. Algorithms that are proposed in the literature use machine learning methods such as decision trees or clustering to differentiate the appliances~\cite{cominola2015benefits}. These algorithms use specific properties of the appliances such as volume, rate, duration, and pressure of the water flow for identification.
%to identify the operational appliances.

Methods for disaggregating natural gas data range from time series disaggregation~\cite{askari2015high} to fuzzy c-means clustering~\cite{fernandes2016analysis} and edge detection~\cite{yamagami1996non}. From aggregated data, these methods can infer the appliance usage and can profile the consumers in terms of their headcount, employment status, social class, and annual income~\cite{fernandes2016analysis}.

\section{Privacy Preservation Techniques for Smart Metered Resources}
\label{section:privacy-techniques}

Since the literature on privacy protection for smart water and gas metering is nonexistent at the moment, in this section, we present the aforementioned privacy preservation methods proposed for HTG electricity metering and make a first attempt on how (and if) they can be adapted for smart water and gas metering.

\subsection{Data Obfuscation}
\label{subsec:obfuscation}
In smart grids, obfuscation methods add noise to the metered consumption data to hide appliance signatures. Some of these methods utilize a rechargeable battery to create noise in the metered data. The charging and discharging times of the battery is scheduled meticulously so that the differences in the power usage levels of different time slots do not leak relevant information.

A similar technique can be used for hiding water and gas usage profiles by employing storage tanks. 
%However, the end-uses of water and gas require a bulk usage of the resource most of the time. Therefore, to ensure privacy, significant storage capacities of water and gas are needed. Also, storing natural gas which is a flammable substance in a household can create safety risks. 
Alternatively, the energy obtained from natural gas can be stored in the form of heated water in a hot water tank to obfuscate both the metered water and metered gas. This can be considered as a more secure way instead of storing a flammable substance in a household.

\subsection{Data Aggregation}
\label{subsec:aggregation}
Aggregation methods combine the metering data of multiple customers and encrypt them, so that the metering information cannot be used by adversaries. Moreover, privacy-enhancing cryptographic methods such as homomorphic encryption are commonly employed for flexibility. Indeed, homomorphic encryption methods enable UCs to do certain calculations on the measured data without needing to decrypt it.

A similar approach can easily be adopted for water and gas metering. Data from electricity, water, and gas meters can also be encrypted together in cases where these services are provided by the same UC. The downside of such methods is that a trusted third party or the UC have to be trusted with the original data.

\subsection{Anonymization}
\label{subsection:anonymization}
Data anonymization methods utilize a pseudonym or an ID that cannot be linked to the consumer rather than the consumer's real-life information to protect the consumer's privacy. However, this does not stop an adversary who has physical access to the feeder cable of a house from obtaining the aggregate data and disaggregating it to infer private information. In fact, with the advancements in \REV{Microelectromechanical Systems (MEMS)} and sensor equipment, measuring water and natural gas from outside a house has become as easy as measuring electricity. Therefore, anonymization can be coupled with other privacy-enhancing methods to be more effective, however, it is not a failproof strategy.

\subsection{Downsampling}
\label{subsection:downsampling}
Smart meters generally send the time averaged value of consumption within each sampling period to the UCs. Therefore, as the sampling frequency of a smart meter decreases, some appliances start and finish their operations within the same sampling period. Since most of the NIALM methods look for edges in the metered data for detecting on/off events, appliances with short operation durations can be missed by the disaggregation algorithms. Therefore, it is safe to state that increasing the sampling period can help privacy preservation. However, sampling less frequently is against the very idea of smart metering. Therefore, although its effect on privacy should always be taken into consideration, downsampling itself is not an acceptable solution for ensuring privacy in smart resource metering.

\subsection{Load Shaping}
\label{subsection:load-shpaing}
Load shaping methods use rechargeable batteries, renewable energy sources, and shiftable appliances to erase the appliance signatures from the metered load as much as possible. The downsides of these methods are the initial monetary costs of household batteries and renewable energy sources in addition to consumer discomfort due to the shifting of appliances' operation windows. The analogue of load shaping for the water and natural gas metering can be achieved by using water tanks as well as by shifting some of the water and gas consuming events (such as hot water and heating), which is not required to be on-demand, to other time slots.

Of all aforementioned privacy preservation techniques, load shaping methods are, arguably, the most promising, as they hide the appliance signatures from any actors from outside the house, while still enabling accurate and frequent meter readings. However, there are a number of appliances that use more than one resource at a time, and all of these resources can be metered in a privacy-threatening frequency. Therefore, no matter which techniques are used, all the metered resources used in a household should be taken into account while considering privacy. In the next section, we define the concept of holistic privacy and provide a case study showing why such an approach is necessary.

\section{Holistic Privacy}
\label{section:holistic-privacy}

\begin{figure}[hbt]
	\centering
    \begin{center}
        \includegraphics[width=0.45\textwidth]{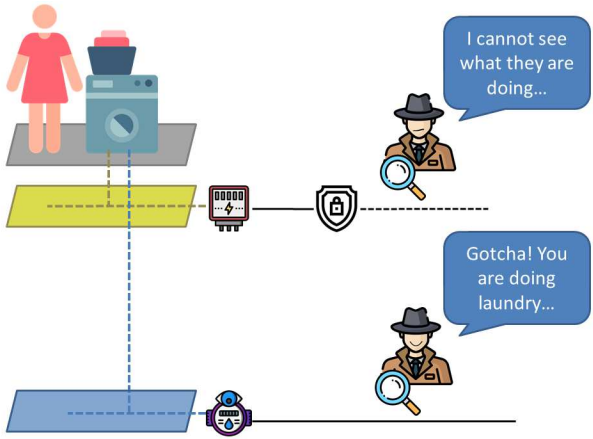}
        \caption{Hiding only electricity data is not sufficient if water and gas are also metered frequently.}
        \label{figure:hiding}
    \end{center}
\end{figure}

Privacy preservation schemes designed for electricity utilization can be adopted to hide electricity, water, and natural gas signatures individually. However, as there is a plurality of appliances which simultaneously consume more than one of these smart metered resources, a holistic privacy notion which incorporates all of the smart metered resources simultaneously is necessary (see Fig.~\ref{figure:hiding}). We call this approach \emph{holistic privacy}.

%Some household appliances that use more than one resources are: combi boilers, HVAC (Heating, Ventilation, and Air Conditioning) systems, dishwashers, washing machines (WMs), and ranges (with gas stove and electric oven). These multiple resource consuming appliances are widely available in a typical household.
Even if the electricity usage data of the household is completely hidden, metered water and gas data can still be used to infer appliance profiles. For example, assume that a load shaping method is employed for hiding only electricity signatures of multiple resource consuming appliances. As the operation schedule of the appliances are changed, their usage of gas and water is also shifted along with their electricity usage because these resources are used interdependently by those appliances. Therefore, when trying to find the optimal privacy preserving scheduling for those appliances, one should consider electricity, water, gas, and any other resource that is subject to smart metering, \emph{holistically}. Appliances that are especially vulnerable to this privacy threat are the ones that use considerable amounts of multiple resources simultaneously. Combi boilers (electricity, water and gas), Heating, Ventilating and Air Conditioning (HVAC) systems (electricity and gas), washing machines (WMs) and dishwashers (electricity and water) are among such appliances.

\begin{figure}[ht]
	\centering
    \begin{center}
        \includegraphics[width=\linewidth]{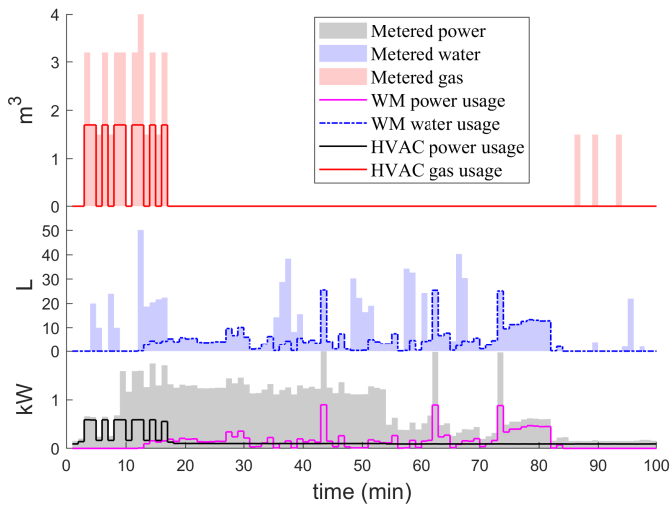}
        \caption{Minutely metered (aggregate) power, water and gas usage of the household along with the individual electric, water, and gas usage of the HVAC and the WM.}
        \label{figure:fig0}
    \end{center}
\end{figure}

Since the research on smart multiple resource metering is relatively new, there are only a few available datasets which include HTG electricity, water, and gas meter data. One useful dataset is the Ampds2 dataset~\cite{makonin2016electricity}, where electricity, water, and natural gas consumption data of a house in Canada from 2012 to 2014 are available. This study also provides consumption data for some of the individual appliances. Therefore, it is possible to observe an individual appliance's power/water/gas signatures on the cumulative load of the household.

For demonstration purposes, we focus on two multiple resource consuming appliances of the house: the HVAC unit and the WM. The HVAC has a furnace that runs on natural gas and an electric furnace fan that blows the hot air into the house. The WM uses both electricity and water for its operation. Sample consumption data of the appliances, as well as the measurement data of the smart electricity (power), water, and natural gas meters of the house are plotted in Fig.~\ref{figure:fig0}. We analyze three privacy preservation cases in the rest of this section.

\subsection{Case 0: No Load Shaping}
\label{subsection:case0}
The information on HVAC and WM usage revealed by the metered power, water, and gas data is obvious to even the naked eye (see Fig.~\ref{figure:fig0}). Also, it can be observed that the electricity and gas usages of the HVAC are highly correlated. Mutual Information (MI) is a metric utilized for measuring information leak in smart metered data in the literature~\cite{kement2017comparative}. In fact, the higher the MI between two time series, the more information they reveal on one another. The MI between HVAC power and metered power is 1.91 bits, whereas the MI between HVAC gas and metered gas is 1.61 bits. Similarly, the correlation between the electricity and water usage of the WM can be observed in Fig.~\ref{figure:fig0}. The MI between WM water usage and metered water data is 0.59 bits, whereas the MI between power usage of the WM and metered power is 0.56 bits. 

\subsection{Case 1: Privacy by Shaping Electricity Only}
\label{subsection:case1}
\begin{figure}[hbt]
	\centering
    \begin{center}
        \includegraphics[width=\linewidth]{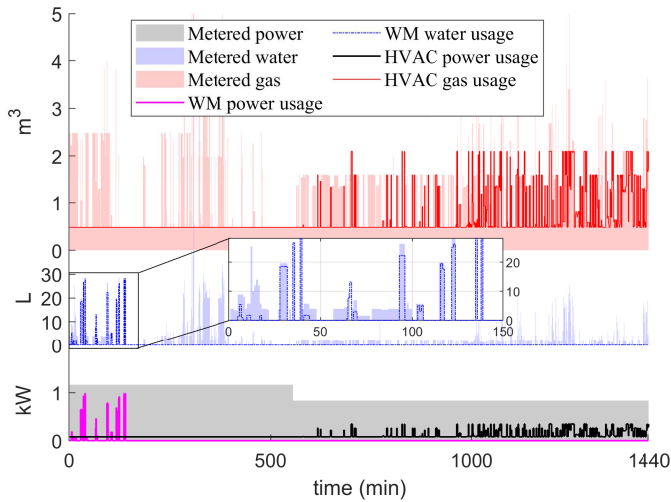}
        \caption{Metered electricity, water, and gas usage along with individual HVAC and WM usages when only electrical signatures are hidden.}
        \label{figure:fig1}
    \end{center}
\end{figure}

We use the mathematical programming framework presented in~\cite{kement2017comparative} to shape the electrical load to hide the appliance signatures. \REV{We utilize the Best Effort (BE) strategy~\cite{kalogridis2010privacy}, which is one of the best performing strategies in terms of privacy~\cite{kement2017comparative}. BE strategy minimizes the difference between metered power usages at adjacent time slots. Under ideal conditions, BE strategy returns a metered load that is flat throughout the day.} Resulting electricity, water, and gas usages of the house are plotted in Fig.~\ref{figure:fig1}. The electricity consumed by HVAC and WM (Fig.~\ref{figure:fig1}, black and purple lines) are completely hidden within the metered power (Fig.~\ref{figure:fig1}, gray area) with the help of a 2~kWh battery assumed in the formulation. The MI of HVAC power and WM power are reduced to 0.23~bits and 0.03~bits, respectively. However, the gas usage of the HVAC and water usage of the WM can still be inferred from the overall metered gas and water data (Fig.~\ref{figure:fig1}, red and blue areas, respectively). 
%\REV{The zoomed part of the water graph in Fig.~\ref{figure:fig1} shows the correlation between the metered water and WM's water usage.} 
\REV{The zoomed section in Fig.~\ref{figure:fig1} reveals the details of the metered water and WM's water usage.} 
The MI between HVAC gas data and metered gas data remains high at 1.82 bits, while the MI between WM water data and metered water data is 0.49 bits.

\subsection{Case 2: Holistic Privacy by Shaping Electricity, Water, and Gas Simultaneously}
\label{subsection:case2}
\begin{figure}[ht]
	\centering
    \begin{center}
        \includegraphics[width=\linewidth]{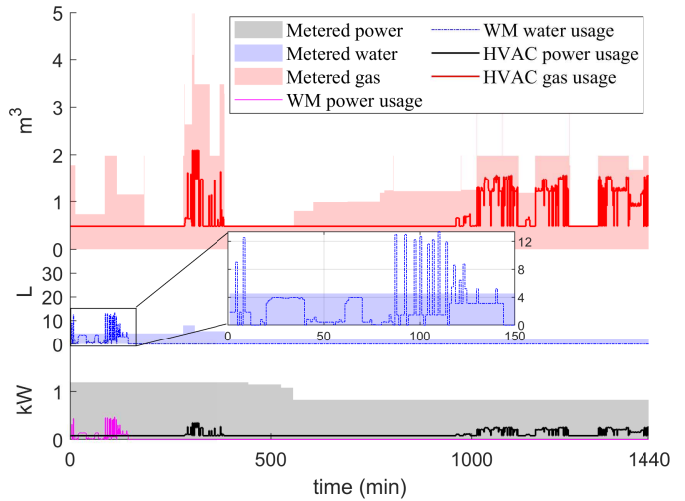}
        \caption{Metered electricity, water, and gas usage along with HVAC and WM signatures with a holistic privacy approach.}
        \label{figure:fig2}
    \end{center}
\end{figure}

We extend the formulation in~\cite{kement2017comparative} to \REV{utilize BE strategy for all three resources and} make all three metered data as flat as possible. We assume there is a water tank (analogous to the household battery) which can be filled/emptied as desired to shape the water demand. We do not assume the existence of a gas tank due to its questionable safety. It can be observed in Fig.~\ref{figure:fig2} that the gas usage of the HVAC and water usage of the WM \REV{(see the zoomed panel in Fig.~\ref{figure:fig2} for a detailed view)} can be hidden more effectively along with their power usage, with a small compromise from the flatness of the metered electricity in comparison to Case 1. The MI values of the HVAC are 0.49 \REV{bits} and 0.58 \REV{bits} for electricity and gas, respectively. For the WM, the MI values are 0.17 and 0.01 bits for electricity and water, respectively.

Note that in Case 2 the electricity is hidden better (with almost no appliance signature) compared to water and gas, which leak more information. The reason for this is the difference in total number of amenities that could be used to shape each resource. In the Ampds2 dataset, there are more than 10 electrical appliances that can be exploited for load shaping as well as a household battery which can be used to flatten the load. In the case of water, there are 4 appliances and a water tank to be employed for load shaping. However, the household has 4 gas consuming appliances, yet, individual data for only one of them (HVAC furnace) is available and there is no gas storage assumed in the household. Thus, the optimal results show a well hidden electrical load compared to water and gas. 

Another interesting point is that shaping all smart metered resources (Case 2) will incur a certain cost and discomfort to the consumer when compared with no load shaping (Case 0). However, the increase in the cost and discomfort in Case 2 is comparable with Case 1, which indicates that the implementation of the holistic privacy strategy is as practical and feasible as the traditional, electricity-only privacy techniques.

The case study we presented demonstrates the effectiveness of the holistic privacy preservation in hiding appliance signatures for all metered resources. As the number of datasets with multiple smart metered resources increases, more detailed analysis can be made not only on the load shaping methods, but on all the other privacy preserving techniques as well.

\section{Conclusion}
\label{section:conclusion}

The current trends and evolution of smart water and gas metering is somewhat similar to the early deployments of smart electricity meters almost a decade ago. Basic operation principles of NIALM methods designed to infer appliance usages from metered aggregate electricity data can also be employed on HTG metered water and natural gas data with certain adaptations which raises privacy concerns for water and gas metering. In this study, we review the advances in smart water and natural gas metering, NIALM techniques for disaggregating these resources, and privacy protection strategies for HTG electricity metering as well as possible adaptations of these methods to HTG water and gas metering. We propose the novel notion of holistic privacy for households that are equipped with HTG electricity, water, and gas meters. We demonstrate the effectiveness of the holistic privacy preserving approach by analyzing the performance of load shaping on the Ampds2 dataset for electricity, water, and gas simultaneously. We hope that this study fosters future research on holistic privacy preservation for the next generation smart homes.

A possible future research direction for the holistic privacy can be altering the Internet traffic. Internet is another resource which is becoming almost as available in households as electricity, water and gas, and it can be used to infer some private information or detect some of the IoT devices. By sniffing the cable modem or the access point, an adversary can deduce private information about the household such as the number of residents (e.g., from Medium Access Control -- MAC -- addresses) and the vacancy of the house (e.g., from Hypertext Transfer Protocol -- HTTP -- traffic). Furthermore, by analyzing the Internet protocols used in the household, some of the smart IoT appliances that use a specific type of such protocols can be identified. Possible countermeasures include using various spoofing techniques to generate fake traffic to hide the real Internet usage and mitigate the effectiveness of adversarial actions. We leave the details of implementing such methods as a future research direction.

\bibliographystyle{IEEEtran}
\bibliography{Main}

\begin{thebibliography}{10}
\providecommand{\url}[1]{#1}
\csname url@samestyle\endcsname
\providecommand{\newblock}{\relax}
\providecommand{\bibinfo}[2]{#2}
\providecommand{\BIBentrySTDinterwordspacing}{\spaceskip=0pt\relax}
\providecommand{\BIBentryALTinterwordstretchfactor}{4}
\providecommand{\BIBentryALTinterwordspacing}{\spaceskip=\fontdimen2\font plus
\BIBentryALTinterwordstretchfactor\fontdimen3\font minus
  \fontdimen4\font\relax}
\providecommand{\BIBforeignlanguage}[2]{{%
\expandafter\ifx\csname l@#1\endcsname\relax
\typeout{** WARNING: IEEEtran.bst: No hyphenation pattern has been}%
\typeout{** loaded for the language `#1'. Using the pattern for}%
\typeout{** the default language instead.}%
\else
\language=\csname l@#1\endcsname
\fi
#2}}
\providecommand{\BIBdecl}{\relax}
\BIBdecl

\bibitem{marmol2012do}
F.~G. {Marmol}, C.~{Sorge}, O.~{Ugus}, and G.~M. {Perez}, ``Do not snoop my
  habits: Preserving privacy in the smart grid,'' \emph{IEEE Commun. Mag.},
  vol.~50, no.~5, pp. 166--172, May 2012.

\bibitem{iot}
\BIBentryALTinterwordspacing
``Smart meter market report 2019-2024,'' IoT Analytics: Market Insights for the
  Internet of Things, Tech. Rep., Nov. 2019. [Online]. Available:
  \url{https://iot-analytics.com/product/smart-meter-market-report-2019-2024/}
\BIBentrySTDinterwordspacing

\bibitem{sun2015comprehensive}
Q.~Sun, H.~Li, Z.~Ma, C.~Wang, J.~Campillo, Q.~Zhang, F.~Wallin, and J.~Guo,
  ``A comprehensive review of smart energy meters in intelligent energy
  networks,'' \emph{IEEE Internet Things J.}, vol.~3, no.~4, pp. 464--479, Aug.
  2015.

\bibitem{kement2017comparative}
C.~E. Kement, H.~Gultekin, B.~Tavli, T.~Girici, and S.~Uludag, ``Comparative
  analysis of load-shaping-based privacy preservation strategies in a smart
  grid,'' \emph{IEEE Trans. Ind. Informat.}, vol.~13, no.~6, pp. 3226--3235,
  Dec. 2017.

\bibitem{mudumbe2015smart}
M.~J. Mudumbe and A.~M. Abu-Mahfouz, ``Smart water meter system for
  user-centric consumption measurement,'' in \emph{Proc. IEEE Int. Conf. Ind.
  Informat. (INDIN)}, 2015, pp. 993--998.

\bibitem{cominola2015benefits}
A.~Cominola, M.~Giuliani, D.~Piga, A.~Castelletti, and A.~E. Rizzoli,
  ``Benefits and challenges of using smart meters for advancing residential
  water demand modeling and management: A review,'' \emph{Environ. Modell.
  Softw.}, vol.~72, pp. 198--214, Oct. 2015.

\bibitem{schantz2014water}
C.~Schantz, J.~Donnal, B.~Sennett, M.~Gillman, S.~Muller, and S.~Leeb, ``Water
  nonintrusive load monitoring,'' \emph{IEEE Sensors J.}, vol.~15, no.~4, pp.
  2177--2185, Apr. 2014.

\bibitem{tewolde2010high}
M.~Tewolde and J.~Longtin, ``High-resolution meter reading system for gas
  utility meter,'' in \emph{Proc. IEEE SENSORS}, 2010, pp. 849--852.

\bibitem{dong2017mems}
S.~Dong, S.~Duan, Q.~Yang, J.~Zhang, G.~Li, and R.~Tao, ``{MEMS}-based smart
  gas metering for internet of things,'' \emph{IEEE Internet Things J.},
  vol.~4, no.~5, pp. 1296--1303, Oct. 2017.

\bibitem{nhede_2018}
\BIBentryALTinterwordspacing
N.~Nhede. (2018, Aug.) Emerging economies to fuel smart gas metering systems
  market expansion. [Online]. Available:
  \url{https://www.smart-energy.com/industry-sectors/smart-meters/gm-insights-smart-gas-metering-systems/}
\BIBentrySTDinterwordspacing

\bibitem{askari2015high}
S.~Askari, N.~Montazerin, and M.~F. Zarandi, ``High-frequency modeling of
  natural gas networks from low-frequency nodal meter readings using
  time-series disaggregation,'' \emph{IEEE Trans. Ind. Informat.}, vol.~12,
  no.~1, pp. 136--147, Feb. 2016.

\bibitem{fernandes2016analysis}
M.~P. Fernandes, J.~L. Viegas, S.~M. Vieira, and J.~M. Sousa, ``Analysis of
  residential natural gas consumers using fuzzy c-means clustering,'' in
  \emph{Proc. IEEE Int. Conf. Fuzzy Syst. (FUZZ-IEEE)}, 2016, pp. 1484--1491.

\bibitem{yamagami1996non}
S.~Yamagami, H.~Nakamura, and A.~Meier, ``Non-intrusive submetering of
  residential gas appliances,'' in \emph{Proc. ACEEE Summer Study on Energy
  Efficiency in Buildings}, 1996, pp. 265--273.

\bibitem{makonin2016electricity}
S.~Makonin, B.~Ellert, I.~V. Bajic, and F.~Popowich, ``Electricity, water, and
  natural gas consumption of a residential house in {Canada} from 2012 to
  2014,'' \emph{Sci. Data}, vol.~3, pp. 160\,037:1--160\,037:12, Jun. 2016.

\bibitem{kalogridis2010privacy}
G.~Kalogridis, C.~Efthymiou, S.~Z. Denic, T.~A. Lewis, and R.~Cepeda, ``Privacy
  for smart meters: Towards undetectable appliance load signatures,'' in
  \emph{Proc. IEEE Int. Conf. Smart Grid Commun. (SmartGridComm)}, 2010, pp.
  232--237.

\end{thebibliography}

\footnotesize
\vspace{0.4cm}
\noindent \textbf{Cihan Emre Kement} [S] (kement@mit.edu) 
is a Ph.D. candidate in the Department of Electrical and Electronics Engineering, TOBB University of Economics and Technology, Ankara, Turkey. He is also a Fulbright visiting researcher at Laboratory for Information and Decision Systems (LIDS), Massachusetts Institute of Technology (MIT), Cambridge, MA, USA.
His current research interests are security and privacy in cyber-physical systems, wireless communications and optimization.

\vspace{0.4cm}
\noindent\textbf{Bulent Tavli} [SM] (btavli@etu.edu.tr) is a professor in the Department of Electrical and Electronics Engineering, TOBB University of Economics and Technology, Ankara, Turkey. His current research areas are network design and analysis, wireless ad hoc and sensor networks, IoT, wireless communications, machine learning for communications systems, optimization, embedded systems, information security and privacy, blockchain, and smart grid.

\vspace{0.4cm}
\noindent\textbf{Hakan Gultekin} (hgultekin@squ.edu.om) is an associate professor in the Department of Mechanical and Industrial Engineering, Sultan Qaboos University, Muscat, Oman and also in the Department of Industrial
Engineering, TOBB University of Economics and Technology, Ankara, Turkey. His research interests include scheduling, optimization modeling, and exact and heuristic algorithm development, especially for problems arising in modern manufacturing systems, energy systems, and wireless sensor networks.

\vspace{0.4cm}
\noindent\textbf{Halim Yanikomeroglu} [F] (halim@sce.carleton.ca) is a full professor in the Department of Systems and
Computer Engineering at Carleton University, Ottawa, Canada. His research interests cover many aspects of 5G/5G+
wireless networks. His collaborative research with industry has resulted in 37 granted patents. He is a Fellow of IEEE,
EIC (Engineering Institute of Canada), and CAE (Canadian Academy of Engineering); he is a Distinguished Speaker for the
IEEE Communications Society and the IEEE Vehicular Technology Society.

\end{document}